\pdfoutput=1

\documentclass[preprint,authoryear]{elsarticle}

\usepackage{natbib}
\setcitestyle{square, comma, numbers}
\usepackage{lineno}
\usepackage[colorlinks]{hyperref}
\usepackage{amsmath,amssymb,amsfonts}
\usepackage{graphicx}
\usepackage{listings}
\usepackage{multirow}
\usepackage{xurl}
\usepackage{pifont}
\usepackage[table,xcdraw]{xcolor}
\usepackage{color, colortbl}
\usepackage{booktabs}
\usepackage{soul}
\usepackage{caption}
\usepackage{subcaption}

\newcommand{\xmark}{\ding{55}}%

\definecolor{LightCyan}{rgb}{0.88,1,1}
\definecolor{Gray}{rgb}{0.83,0.83,0.83}
\definecolor{White}{rgb}{1,1,1}
\newcolumntype{a}{>{\columncolor{Gray}}c}

\biboptions{authoryear}

\begin{document}

\begin{frontmatter}

\title{Negligible effect of brain MRI data preprocessing for tumor segmentation.}

\author[skoltech,airi]{Ekaterina Kondrateva\corref{mycorrespondingauthor}}
\author[skoltech,airi]{Polina Druzhinina\corref{mycorrespondingauthor}}
\author[burdenko]{Alexandra Dalechina}
\author[burdenko]{Svetlana Zolotova}
\author[burdenko]{Andrey Golanov}
\author[skoltech,airi]{Boris Shirokikh}
\author[skoltech,airi]{Mikhail Belyaev}
\author[airi]{Anvar Kurmukov\corref{mycorrespondingauthor}}
\cortext[mycorrespondingauthor]{Authors contributed equally}

\address[airi]{Artificial Intelligence Research Institute (AIRI), Moscow, Russia}
\address[skoltech]{Skolkovo Institute of Science and Technology, Moscow, Russia}
\address[burdenko]{National Medical Research Center for Neurosurgery, Moscow Gamma Knife Center, Moscow, Russia}

\begin{abstract}

Magnetic resonance imaging (MRI) data is heterogeneous due to differences in device manufacturers, scanning protocols, and inter-subject variability. A conventional way to mitigate MR image heterogeneity is to apply preprocessing transformations such as anatomy alignment, voxel resampling, signal intensity
equalization, image denoising, and localization of regions of interest. Although a preprocessing pipeline standardizes image appearance, its influence on the quality of image segmentation and on other downstream tasks in deep neural networks has never been rigorously studied.

Experiments on three publicly available datasets evaluate
the effect of different preprocessing steps in intra- and inter-dataset training scenarios. Results demonstrate that most popular standardization steps add no value to network performance; moreover, preprocessing can hamper performance. Our results suggest that image intensity normalization approaches do not contribute to model accuracy because of the reduction of signal variance with image standardization. Additionally, the contribution of skull-stripping in data preprocessing is almost negligible if measured in terms of estimated tumor volume.

The only essential transformation for accurate deep learning
analysis is the unification of voxel spacing across the dataset. In contrast, inter-subjects anatomy alignment in the form of non-rigid atlas registration is not necessary and intensity equalization steps (denoising, bias-field correction and
histogram matching) do not improve performance. 

The study code is accessible online\footnote{https://github.com/MedImAIR/brain-mri-processing-pipeline}.

\end{abstract}


\end{frontmatter}


\section{Introduction}
In recent years, modern deep neural networks (DNN) have steadily improved the quality of automatic segmentation pipelines in medical imaging. Specifically, for the task of brain tumor segmentation, the performance of DNNs has achieved human-level efficiency \cite{chang2019automatic}. This advancement can be explained by the improvement of DNN architectures and the growth of the training datasets. For example, the size of the BraTS competition \cite{bakas2018identifying} training dataset increased from $100$ subjects in 2013 to $2000$ subjects in 2021. Simultaneously, top-performing algorithms have progressed from random forest and gradient boosting trees on radiomics features: first to fully convolutional networks, then to u-shaped UNet and UNet-like networks, and finally to Vision transformers. 

In contrast, the preprocessing steps used to prepare data for analysis seem to have undergone considerably fewer changes. For instance, the set of preprocessing steps for brain MRI images has remained relatively stable and has been reproduced across the majority of papers on the topic from the early 2010s until now. Here we challenge the conventional pipelines for MRI image processing and question their necessity for accurate prediction with regard to new advanced deep learning machinery.

The traditional brain MRI preparation steps could be divided into four distinct categories: 
\begin{itemize}
    \item The first category is subject-wise image alignment, typically in the form of rigid registration of one MRI sequence to another, (e.g. T2-FLAIR onto T1 with contrast enhancement). This step is mandatory if one uses multiple MR modalities to predict a single segmentation map to ensure correct alignment between ground truth annotation and corresponding image's voxels. 
    \item The second category is voxels resampling to some standard. The most common methods are voxel resampling to homogeneous spacing (often $1\times 1\times 1\text{ mm}^3$) and non-rigid registration to some anatomical atlas. 
    \item The third category includes steps that affect voxels' intensity distribution, such as bias-field correction (such as N4 correction \cite{tustison2010n4itk}), intensity normalization (typically in a form of image-wise z-scoring), image denoising methods (e.g. SUSAN \cite{smith1997susan}), and histogram equalization \cite{nyul2000new}.
    \item  Finally, the last step that is preserved in almost all the papers is skull stripping as a method to localize regions of interest (the brain tissue), implement feature selection to ease localization, and reduce the amount of False Positives \cite{chang2019automatic}.
\end{itemize}

While the motivation behind applying these steps is clearly to standardize image appearance and remove different sources of domain shift \cite{kondrateva2021domain}, these steps are computationally costly and their utility for deep-learning segmentation lacks investigation. Specifically, it is widely known that increasing variability of the data by data augmentation (image resizing, non linear intensity transformations, applying noise, etc.) leads to improved DNN performance \cite{wightman2021resnet}. However, data preprocessing works quite in the opposite way by reducing data variance.

In this study we analyze the most popular preprocessing steps for brain MRIs and measure their influence on deep-learning based tumor segmentation tasks. We analyze different preprocessing strategies and recommend the minimal pipeline required for accurate segmentation with the benefits of lower computational costs and improved reproducibility.


\section{Related works}

Image preprocessing is a de-facto standard first step in almost all deep learning pipelines for medical image analysis \cite{nixon2019feature}. In this domain, data preparation is convenient due to two major causes:
\begin{itemize}
    \item diversity in scanning protocols and therefore diverse spatial resolutions and image intensity profiles \cite{kurmukov2021challenges},
    \item large image sizes and small sample sizes, thus leading to high-dimension learning compounding negative effects on model generalisability \cite{berisha2021digital}.
\end{itemize} 

For example, a typical multi-institutional brain MRI dataset consists of images with varying resolutions and acquisition parameters (depending on scanning protocol). Therefore, a majority of studies utilize data preprocessing pipelines. We select several recent publications on brain MRI segmentation to identify the most common preprocessing steps (see Table \ref{tab-preprocessing-papers}).

\begin{table}[h!]
\caption{Common preprocessing steps for multi-modal brain MRI image analysis. Checkmarks represent the step mentioned in the study and x-marks are placed if the step is missing or unclear.}
\label{tab-preprocessing-papers}
\centering
\resizebox{\textwidth}{!}{
\begin{tabular}{lcccccccccc}
    \toprule
    Preprocessing step & \multirow{2}{6em}{Resample to image size} & \multirow{2}{6em}{Resample to spacing} & \multirow{2}{6em}{Atlas registration} & \multirow{2}{6em}{Bias-field correction} & Denoising & \multirow{2}{6em}{Histogram matching} & \multirow{2}{6em}{Skull stripping} \\
    \\
    \midrule
    \cite{gyHorfi2021fully} &\checkmark&\checkmark&\checkmark&\checkmark&\xmark&\checkmark&\checkmark\\
    \cite{ranjbarzadeh2021brain}&\checkmark&\checkmark&\checkmark&\xmark&\xmark&\xmark&\checkmark\\
    \cite{pei2020longitudinal}&\checkmark&\checkmark&\checkmark&\checkmark&\checkmark&\checkmark&\checkmark\\
    \cite{menze2021analyzing}&\xmark &\checkmark&\xmark&\xmark&\xmark&\xmark&\checkmark\\
    \cite{ermics2020fully}&\checkmark&\xmark&\xmark&\xmark&\xmark&\xmark&\checkmark \\
    \cite{eijgelaar2020robust}&\checkmark&\checkmark&\checkmark&\checkmark&\xmark&\xmark&\checkmark \\
    \cite{rathore2017brain}&\checkmark&\checkmark&\checkmark&\checkmark&\checkmark&\checkmark&\checkmark \\
    \cite{wang2019automatic}&\xmark&\xmark&\xmark&\xmark&\xmark&\xmark&\xmark \\
    \cite{bakas2018identifying}&\xmark&\xmark&\checkmark&\xmark&\xmark&\xmark&\checkmark \\
    \cite{kickingereder2019automated}&\xmark&\xmark&\xmark&\xmark&\xmark&\xmark&\checkmark \\
    \cite{bakas2022university}&\xmark&\xmark&\checkmark&\xmark&\xmark&\xmark&\checkmark\\
    \bottomrule
\end{tabular}}
\end{table}

The overall brain MRI data preprocessing pipeline can be divided into four groups of methods.

\subsection{Intra-subject alignment (rigid registration)} During this step different MR sequences from a single patient are reoriented in a similar way and rigidly registered. This step is applied in all studies that analyze multi-sequence MRI \cite{gyHorfi2021fully,ranjbarzadeh2021brain,pei2020longitudinal,menze2021analyzing,ermics2020fully,eijgelaar2020robust}.

\subsection{Inter-subject alignment} This step standardizes the size of images across the dataset. Most of the observed papers use voxel resampling to an isotropic voxel (e.g. $1 \times 1 \times 1 \text{ mm}^3$), or to the same image resolution (in voxels, e.g. $256 \times 256 \times 256$), or both, by means of non-rigid atlas registration \cite{gyHorfi2021fully,ranjbarzadeh2021brain,pei2020longitudinal,menze2021analyzing} and \cite{ermics2020fully,eijgelaar2020robust,rathore2017brain,bakas2018identifying}. The only exceptions are two studies that analyze the data acquired with a unified scanning protocol (isotropic voxel) \cite{kickingereder2019automated,wang2019automatic}.

\subsection{Non-linear intensity correction and enhancement}  Multi-institutional studies typically use some intensity or noise correction approaches, with the most popular being histogram matching \cite{gyHorfi2021fully,pei2020longitudinal,rathore2017brain}, bias-field correction \cite{gyHorfi2021fully,pei2020longitudinal,eijgelaar2020robust,rathore2017brain}, and denoising \cite{pei2020longitudinal,rathore2017brain}.

\begin{figure}[ht]
\includegraphics[width=\textwidth]{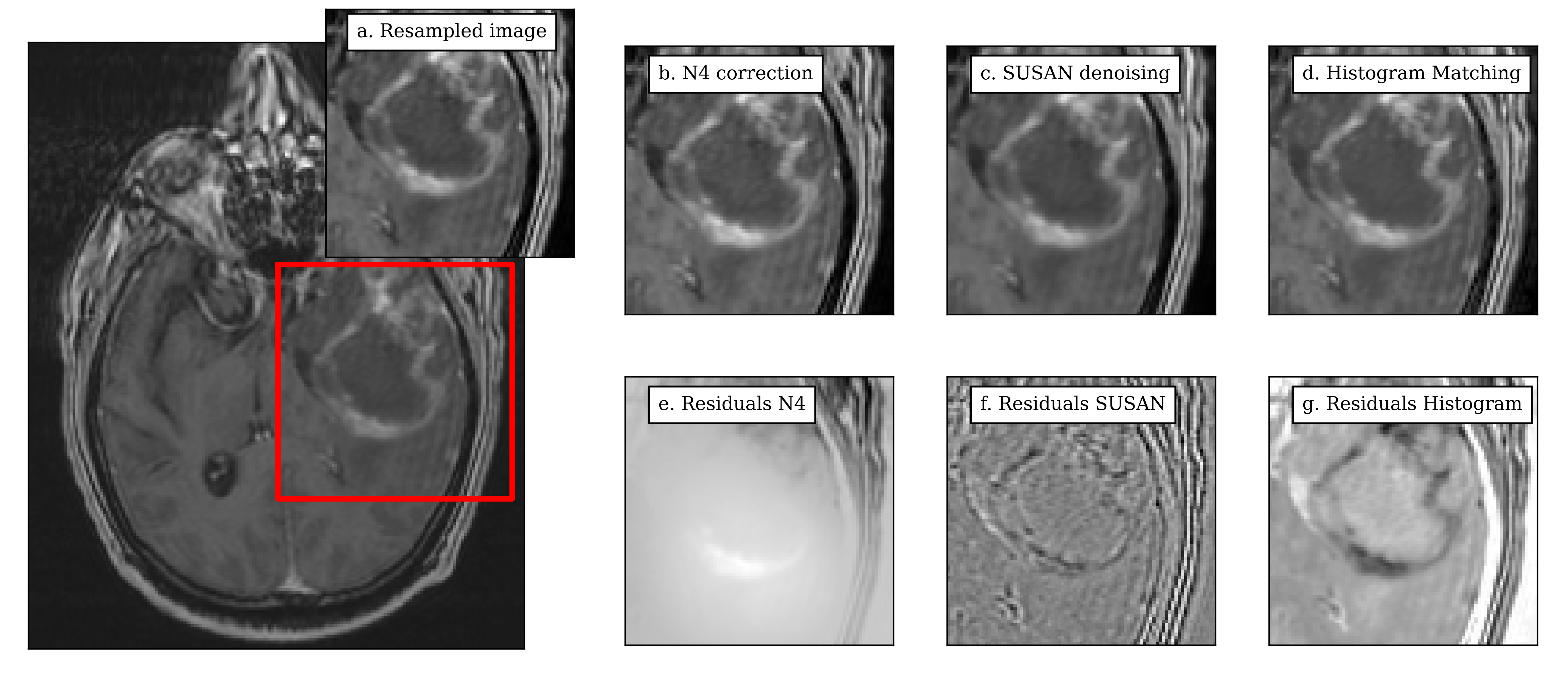}
\caption{Visualization of three methods of image intensity correction used in the ablations study. The top raw (b-d) showing the image segment with tumor after preprocessing of (a) an example CT1 MR image from GBM data after resampling to [1,1,1] voxel size; (b) N4 correction of the image; (c) SUSAN denoising of the image; (d) Histogram standartization to GBM sample; The bottom row (e-g) showing image residuals as voxelwise difference between the original resampled image and the preprocessed one.}
\label{fig-image-correction}
\end{figure}

Histogram equalization (harmonization or matching) methods standardize images by aligning their intensity distributions \cite{nyul2000new}. Denoising algorithms filter the image whilst preserving the underlying structure \cite{smith1997susan}. Finally, bias-field correction methods mitigate the effects of magnetic field variation \cite{tustison2010n4itk}, see Figure \ref{fig-image-correction}.

Most of the observed studies apply z-scoring \cite{ranjbarzadeh2021brain,pei2020longitudinal,menze2021analyzing,ermics2020fully,rathore2017brain,wang2019automatic,kickingereder2019automated} prior to analysis. This is a common data normalization approach for all computer vision algorithms \cite{patro2015normalization} and is not specific to medical imaging.

\subsection{Skull stripping} Finally, all but one of the observed papers \cite{wang2019automatic} use skull stripping, arguing that  non-brain tissue is a significant source of error for downstream tumor segmentation \cite{chang2019automatic}. Authors point out that skull stripping reduces the number of False Positives and improves segmentation quality. 


\subsection{Novelty of the proposed study}
In general, researchers experimenting on single-institutional data or data collected under unified acquisition protocol tend to use fewer preprocessing steps. On the contrary, the analysis of heterogeneous multi-center data typically includes a data preparation pipeline. For example, the data preparation pipeline for the most-known benchmark dataset for multi-institutional brain MRI segmentation\footnote{https://cbica.github.io/CaPTk/preprocessing\_brats.html} includes image reorientation, altas registration \cite{rohlfing2010sri24}, bias-field correction, and skull stripping \cite{thakur2020brain}.

These MRI preprocessing methods have been exploited by scientists for several decades. Yet to date there is no consensus regarding which of the methods should be applied for deep-learning-based analysis. In the present work we demonstrate that the effect of most standardization steps is negligible even for relatively small data collection. 

To the best of our knowledge, there have been two similar attempts to analyze the influence of preprocessing steps for medical imaging tasks. Authors of \cite{moradmand2020impact} test how preprocessing influences radiomics features calculation (for brain MRI), and \cite{de2021effect} investigate the influence of data augmentation for three medical imaging tasks (brain and knee MRI and liver CT segmentation).

\subsection{Contributions}
In our study we focus on preprocessing steps used primarily for brain MRI. Our contribution is three-fold. First, we numerically estimate the effect of 7 common preprocessing steps: resizing images to equal size,inter-subject atlas registration,resampling voxels to equal size, bias-field correction, image denoising, histogram matching, skull stripping; with state-of-the-art deep learning architectures that are both convolutional and attention-based. This effect is assessed in two scenarios: training from scratch and fine-tuning from a larger dataset.

Second, we propose an explanation for the observed low (or even negative) influence of image intensity correction techniques on model accuracy. For that we compare segmented tissue with the rest of the brain before and after preprocessing.

Third, we suggest a minimal preprocessing pipeline for multi-sequence, multi-protocol brain MRI segmentation studies, consisting of two steps\footnote{We assume intra-subject sequence alignment and z-scoring to be mandatory first and last preprocessing steps, respectively.}:
\begin{enumerate}
    \item voxel resampling (to roughly align images across the dataset),
    \item skull stripping.
\end{enumerate} 
The last step results in a marginal but statistically significant improvement in terms of segmentation quality, especially for smaller ($<100$ subjects) datasets, though it could be omitted for larger datasets with little drop in accuracy.



\section{Methods}

\begin{figure}
\includegraphics[width=\textwidth]{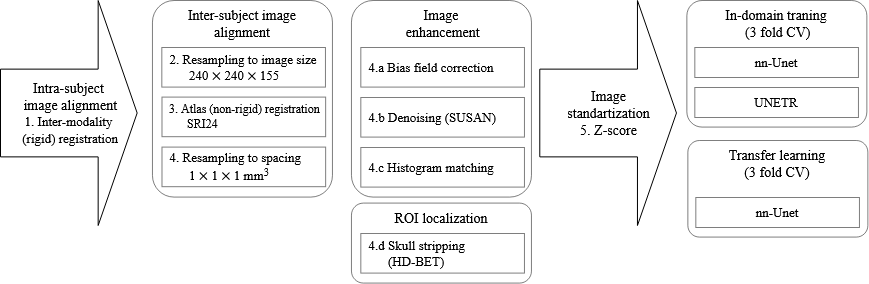}
\caption{Study experimental pipeline. Steps in arrows are mandatory, steps in blocks are optional. Steps 4.a-d are performed after step 4. The minimal preprocessing pipeline consists of steps 1 and 5. Thus, we have 8 different preprocessing pipelines in total.}
\label{fig-pipeline}
\end{figure}

In the current study we test how different preprocessing pipelines affect the quality of a downstream deep learning segmentation model. We compare eight preprocessing pipelines using two segmentation models (convolutional and attention based), Figure \ref{fig-pipeline}. Additionally, we test if the optimal preprocessing pipeline changes in a transfer learning scenario (vs training from scratch) by fine-tuning a model pretrained on a different dataset (with the same preprocessing).

In the following sections we provide information on experimental design and a detailed description of the preprocessing steps, neural network setups, quality metrics, and datasets. 

\subsection{Experimental design} 
We compare preprocessing pipelines  (Figure \ref{fig-pipeline}) in two scenarios:

\begin{enumerate}
    \item Training a segmentation model from scratch.
    \item Fine-tuning a segmentation model from a related or a similar task. 
\end{enumerate}

Thus, we check if different preprocessing pipelines result in better in-domain training, and an improved fine-tuning/knowledge transfer in a cross-dataset regime. In the second scenario we only consider whole model fine-tuning as it is the most common approach to knowledge transfer \cite{wang2021u,dai2019transfer}.

\subsection{Preprocessing steps}
In all experiemnts we use multi-sequence datasets consisting of 4 MRI sequences: T1 weighted (T1), T1 weighted with contrast enhancement (CT1), T2 weighted (T2), and T2-FLAIR (FLAIR).
We start each preprocessing pipeline with rigid registration (rotation, shift and linear interpolation of voxel size) of every MR sequence to CT1 or FLAIR (depending on the dataset) independently for each subject. This step aligns different MR sequences (from the same subject) with each other and the existing tumour annotation. After that, we proceed with preprocessing steps as described in Figure \ref{fig-pipeline}. We end each preprocessing pipeline with image-wise Z-scoring: 
$$
X_s = \frac{X - \text{mean}(X)}{\text{std} (X)}.
$$ 

\subsubsection{Inter-subject alignment} 
After the intra-subject rigid registration, we apply one of three methods to align images across the dataset (inter-subject):
 \begin{itemize}
     \item resizing images to the same  size $240\times 240\times 155$ voxels;
     \item resampling to an isotropic voxels' size $1\times1\times1 \ \text{mm}^3$;
     \item a non-rigid atlas registration to SRI24 atlas \cite{rohlfing2010sri24}. 
 \end{itemize}
 The latter results in both of the same images of size $240\times 240\times 155$, and an isotropic voxel $1\times1\times1 \ \text{mm}^3$. All transformations were performed by ANTs utilities \cite{avants2009advanced}.

\subsubsection{Image enhancement} 
The next step is applying one of three algorithms of image intensity correction: 
\begin{itemize}
    \item bias-field correction,
    \item denoising,
    \item histogram standardization.
\end{itemize}   

Bias-field is a smooth, non-heterogeneous and low-frequency signal  which corrupts MRI images. It is generally believed that algorithms which postprocess MRI images such as segmentation or classification do not produce satisfactory results without correcting for the bias field signal \cite{juntu2005bias}. To test the impact of such a correction we use a popular N4 algorithm \cite{tustison2010n4itk}, implemented in Simple-ITK \cite{beare2018image} as a containerized solution from the CaPTk toolbox. 

 SUSAN is a method of  nonlinear image smoothing and denoising \cite{smith1997susan} that, while old, is still used nowadays for noise reduction. We use an FSL implementation of SUSAN. These two steps were applied for each MR image individually.

Finally, to homogenize grayvalue distribution across images in the dataset, we apply a histogram equalization \cite{nyul2000new}, as implemented in TorchIO library \cite{perez-garcia_torchio_2021}. To estimate the parameters of histogram matching we use training folds, and the same transformation is applied to the data from test folds. Histogram matching is applied sequence-wise, meaning that we equalize voxels' histograms for each MR sequence separately \cite{li2021impact}.

These steps are applied after resampling images to an isotropic voxel (step 4, Figure \ref{fig-pipeline}). As we show in the results, there is no statistical differences between steps 2-4. We thus use voxel resampling as the most intuitive and simple one.

\subsubsection{Skull stripping}
Skull stripping is a method to localize the region of interest and filter out potential False Positives. To test how skull stripping affects model performance, we use HD-BET \cite{isensee2019automated}, which is the most accurate publicly available brain extraction tool. We extract brain mask on CT1 and apply it to all other MR modalities after image alignment (this results in no loss of tumor mask).

\subsection{Models architecture and training}

We test the effect of preprocessing on two deep-learning segmentation architectures: 3D nn-UNet \cite{isensee2021nnu} and vision transformer-based 3D UNETR \cite{Hatamizadeh2022UNETRTF}.

In all experiments we trained networks for 300 epochs or until convergence (using 20 epochs patience as a stopping criterion) without data augmentations, as data augmentation can interfere with measurements of preprocessing effects. All experiments were performed on a 3-fold cross-validation (subject-wise) with the average time for  one experiment being 20 hours on 32 GB Tesla V100.

\subsection{Perfomance metrics}
The accuracy of brain MRI semantic segmentation is conventionally evaluated using the Dice Similarity coefficient (referred to as Dice) and the 95th percentile of the Hausdorff distance between the predicted labels and the provided ground truth, these metrics are used for SOTA models evaluation on benchmark datasets. \cite{bakas2022university}.

In the current work we focus on overlap-based metric Dice coefficient as the standard for assessing the quality of segmentation and provide clear and meaningful interpretations:
\begin{equation}
\text{Dice}\ (A, B) = \frac{2\cdot |A \cap B|}{|A|+|B|},    
\end{equation}
where $A$ and $B$ are 3D binary arrays. In addition, to measure the error in terms of tumor volume estimation, which is the simplest clinically relevant feature \cite{chang2019automatic}, we use Mean Absolute Error (MAE):
\begin{equation}
    \text{MAE}(V_{\text{true}}, V_{\text{estimated}}) = \frac{1}{n}\sum_{i=1}^{n}|V^i_{\text{true}} - V^i_{\text{estimated}}|.
\end{equation}
Mean and standard deviation for Dice score and absolute errors were obtained out of fold. We also asses absolute volumetric error of estimation (for evaluation of result significance from the clinical perspective). 

To measure the differences in voxel intensity between healthy and tumored tissues for different image enhancement experiments, we use Kullback-Leibler divergence (KL) between corresponding intensity histograms:
\begin{equation}
    \text{KL}(H_{\text{healthy}}, H_{\text{tumor}}) = \sum_{i=1}^{\text{\#bins}}H^i_{\text{healthy}}\cdot \log{\frac{H^i_{\text{healthy}}}{H^i_{\text{tumor}}}}
\end{equation}
We report KL values for histograms with fixed
bin size (100 bins), and we have tested the Freedman-Diaconis heuristic \cite{freedman1981histogram} (500-700 bins depending on subject) and Sturge’s rule \cite{sturges1926choice} (20-60 bins). All approaches result
in different absolute KL values but preserve a relative trend, without affecting the experiment's conclusion. 

To assess the results' significance and compare experiments with different data preprocessing, we test the Null hypothesis of the Means equality in two samples with Welch’s t-test Bonferroni corrected for multiple comparisons. If the $p_{\text{value}}$ for the test has not exceeded $0.05$, we declare that there is no evidence in the data for rejecting the Null hypothesis. For simplicity, from now on we will refer to it as statistically significant.

\subsection{Data description}

We explore the effect of data preprocessing for tumor segmentation on multi-modal brain MRI data (a task similar to an extensively studied BraTS \cite{menze2014multimodal}). We selected three largest publicly available multi-domain with original DICOM data from TCIA \cite{Clark2013}:
Glioblastoma Multiform (GBM) and Lower Grade Glioma (LGG) \cite{beers2018dicom, Bakas2017, beers2018dicomseg}, and an in-house dataset of 180 patients with glioblastoma (BGPD)\cite{zolotova2023burdenko}. 

All three datasets contain 4 MR sequences (T1, T2, CT1, FLAIR) available in a raw DICOM format, without any prior preprocessing. All three datasets comprise of multi-protocol studies (Table \ref{tab-datasets-aquis}). A summary of all datasets is presented in Table \ref{tab-datasets}. A more detailed description is given below.

\begin{table}[ht]
\centering
\caption{\label{tab-datasets} Description of open-source multi-institutional datasets on brain tumor segmentation used in the study. All datasets have multimodal MR set per patient, including: CT1, T1, T2 and FLAIR modalities.}
\resizebox{\textwidth}{!}{
\begin{tabular}{ccccccc}
    \toprule
    Dataset & Dataset name  & Size & Diagnosis & Preprocessing & Segmentation classes & Annotation source\\
    \midrule
    
    \cite{beers2018dicomseg} & LGG & 38 & pre-operative low-grade glioma& \xmark & WT, ET & semi-automatic\\
    
    \cite{beers2018dicomseg} & GBM & 102 &pre-operative glioblastoma & \xmark & WT, ET, TC & semi-automatic\\ 
    
    \cite{zolotova2023burdenko}& BGPD & 180 & pre-radiotherapy glioblastoma & \xmark & GTV & manual \\
    
    
    \bottomrule
\end{tabular}}
\end{table}

\subsubsection{Glioblastoma Multiforme (GBM) Dataset}
GBM is an open-source data collection \cite{bakas2017segmentation,bakas2017advancing,clark2013cancer,beers2018dicom} which originally  included 262 patients with preprocessed images. We selected 102 patients with accessible segmentation labels for data without preprocessing\footnote{https://wiki.cancerimagingarchive.net/pages/viewpage.action?pageId=41517733}. The annotations are semi-automatic, obtained using a GLISTRBoost \cite{bakas2015glistrboost} tool with manual correction, and include segmentation for the whole tumor (WT), tumor core (TC) and enhancing tumor (ET). The data subset is comprised of data from $3$ manufacturers and at least $28$ different study protocols. 

\subsubsection{Low Grade Glioma (LGG) Dataset}
LGG is an open-source data collection \cite{pedano2016cancer} that originally included 199 patients. We selected a subset of 38 unique patients with accessible segmentation labels for data without  preprocessing. The annotations are semi-automatic, obtained using a GLISTRBoost \cite{bakas2015glistrboost} tool with manual correction, and include segmentation for WT and ET ($38$ out of the selected $39$ images do not contain TC segmentation map and are thus excluded from the experiments for LGG). The ata subset is comprised  of data from $3$ manufacturers and at least $11$ different study protocols.

\subsubsection{Burdenko Glioblastomas Progression Dataset (BGPD)} 
BGPD \cite{zolotova2023burdenko} is an MRI collection of patients who underwent radiotherapy in Burdenko Neurosurgery Institute for Radiotherapy in Moscow, Russia. It contains the data of 180 unique patients. Segmentation maps were imported from a radiotherapy planning system and correspond to Gross Tumor Volume (GTV). The collection is highly heterogeneous and comprised of data from $4$ manufacturers and at least $51$ different study protocols. 

\begin{table}[ht]
\centering
\caption{\label{tab-datasets-aquis} Variability of study protocols for T1 and T2 MRI sequences for GBM, BGPD and LGG data collections. All datasets contain images from 3 major MRI scanner manufacturers: GE, Siemens, and Toshiba.}
\resizebox{\textwidth}{!}{
\begin{tabular}{cccccccc} 
\toprule
\multicolumn{2}{c}{\multirow{2}{12em}{Aquisition parameter}} & \multicolumn{2}{c}{GBM} & \multicolumn{2}{c}{BGPD} & \multicolumn{2}{c}{LGG} \\
\cmidrule(lr){3-4}\cmidrule(lr){5-6}\cmidrule(lr){7-8}
 & & T1 & T2 & T1 & T2 & T1 & T2\\
\midrule
\multirow{3}{9em}{Echo Time, ms} & min & $2.1$ & $20$ & $1.8$ & $18.4$ & $3.7$& $16.1$\\ 
& max & $19$ & $120$ & $23$ & $120$ & $15$&$120$\\ 
& $\#\text{unique}$ & $28$ & $38$ & $51$ & $67$ &$11$ &$17$\\ 
\midrule
\multirow{3}{9em}{Repetition Time, ms} & min & $5$ & $2020$ & $7.4$&$567$&$8$ &$897$\\ 
& max & $3379.6$ & $6650$ & $3119.2$& $8200$&$3232$ &$10000$\\ 
& $\#\text{unique}$ & $56$ & $36$ & $50$&$57$&$38$ & $18$\\ 
\midrule
\multirow{3}{9em}{Voxel volume, $\text{mm}^3$} & min & $0.5$ & $0.2$ & $0.1$&$0.1$&$0.6$ &$0.5$\\ 
& max & $5.2$ & $5.2$ & $5.3$& $4.8$ & $13.2$&$35.2$\\ 
& $\#\text{unique}$ & $32$ & $32$ & $53$&$60$ &$ 17$&$19$\\ 
\bottomrule
\end{tabular}}
\end{table}

\section{Results}
\begingroup
\setlength{\tabcolsep}{4pt} 
\begin{table}
\centering
\caption{\label{tab-dices} nnU-net and UNETR segmentation performance for the three datasets: GBM and LGG (WT label), BGPD (GTV label). Segmentation accuracy presented in Dice scores from three fold cross-validation as $\text{Mean} \ (\text{STD})$ multiplied by 100, the higher — the better. Models trained for 300 epochs. Arrows denote the statistically significant difference ($p_{\text{value}} < 0.05$) \text{compared to step 4. Resampling to spacing}, $\uparrow$ (increase), $\downarrow$ (decrease).}
\resizebox{\textwidth}{!}{
    \begin{tabular}{lllllll}
        \toprule
        \multicolumn{1}{c}{} & \multicolumn{3}{c}{nnU-net} & \multicolumn{3}{c}{UNETR} \\
        \cmidrule(lr){2-4}\cmidrule(lr){5-7}
        Data preprocessing              & {GBM}     & {BGPD}      & {LGG}      & {GBM}     & {BGPD}    & {LGG} \\ 
        \toprule
        1. Inter-modality registration &  $44\ (28)\downarrow$ & $36\ (29)\downarrow$ & $67\ (27)$ &$39\ (26)\downarrow$ &$35\ (30)\downarrow$ & $66\ (23)$\\ 
        \toprule
        2. Resampling to image size&      $85\ (11)$ & $73\ (19)$ & $72\ (24)$ &$82\ (12)$ &$67\ (20)$ &$66\ (26)$\\
        3. Atlas registration&            $85\ (11)$ & $75\ (16)$ & $71\ (25)$ &$82\ (13)$ &$68\ (21)$ &$67\ (25)$\\  
        \rowcolor{Gray}
        4. Resampling to spacing&         $85\ (12)$ & $74\ (18)$ & $70\ (25)$ &$83\ (14)$ &$67\ (21)$ &$67\ (23)$\\ 
        \midrule
        4.a Bias field correction&        $82\ (13)\downarrow$ & $75\ (17)$ & $67\ (25)$ &$80\ (13)$ &72$\ (19)\uparrow$ &$62\ (22)$\\
        4.b Denoising&                    $84\ (12)$ & $74\ (17)$ &  $70\ (26)$ & $83\ (13)$ &$69\ (21)\uparrow$ &$65\ (25)$\\
        4.c Histogram matching&           $83\ (16)$ & $75\ (16)$ & $68\ (26)$ & $81\ (16)$ &$68\ (18)$ &$63\ (26)$\\ 
        4.d Skull stripping&              $87\ (11)$ & $76\ (14)\uparrow$ & $77\ (21)\uparrow$ &$85\ (11)$ &$72\ (18)\uparrow$ &$75\ (19)\uparrow$\\
        \bottomrule
    \end{tabular}}
\end{table}
\endgroup

Our results are four-fold. First, we analyze the effect of image resampling (Steps 2-4 Figure \ref{fig-pipeline}). Second, we analyze image enhancement methods (Steps 4.a-4.c Figure \ref{fig-pipeline}). Next, we discuss the utility of skull-stripping in terms of segmentation metrics and volume estimates. Finally, we analyze if there is optimal preprocessing in a transfer learning scenario. We end the results section with our recommendations on MRI images preprocessing for deep learning-based tumor segmentation. 

\subsection{Inter-subject image alignment} 
Table \ref{tab-dices} shows validation results of nnU-net and UNETR architectures for the three datasets. 

First, for two larger datasets (GBM and BGPD) we observe that introducing some resampling strategy to homogenise voxel volume across the dataset is always beneficial, Table \ref{tab-dices} step 1 versus steps 2-4. Recall that without any voxel resampling, the differences between voxels' volumes are as large as 10 times for GBM ($0.5\ \text{mm}^3$ versus $5.2\ \text{mm}^3$), and 53 times for BGPD ($0.1\ \text{mm}^3$ versus $5.3\ \text{mm}^3$), see Table \ref{tab-datasets-aquis}. Thus, for a 3D convolutional neural network, the receptive field of a convolution filter will differ by a factor of 53. Interestingly, while for the LGG dataset we also have a difference between voxels' spacing by a factor of 22, there are no significant differences in performance between steps 1-4. The latter might be the consequence of a relatively small sample size.

Second, we observe that applying non-rigid Atlas registration (step 3) lead to the same results compared to a faster Resampling to the same image size (step 2) or Resampling all images to the same voxel spacing (step 4). For both NN architectures in all datasets, it is not possible to reject the Null  hypothesis about the equality of means ($p_{\text{values}} > 0.1$, accounting multiple comparisons correction). Recall that images on step 3 are both the same image size and have a voxel of the same volume\footnote{Resampling to same image size results in almost equal voxel volumes  $1.07\ (0.34) \ \text{mm}^3$, mean (std) for BGPD dataset.}. 

\subsection{Image enhancement}

We compare three different intensity normalization steps commonly used in brain MRI analysis pipelines: Bias field correction (step 4.a), Denoising (step 4.b) and Histogram matching (step 4.c). As there were no significant differences found between resampling approaches, steps 4.a-c were performed after resampling images to the same voxel volume (step 4). 

First, for a convolutional nnU-net, intensity correction transformations could be completely omitted. As show in Table \ref{tab-dices}, there is no statistically significant improvement of either of steps 4.a-c compared to step 4 ($p_{\text{values}} > 0.1$) for all three datasets. In most of the cases, the average segmentation performance is actually worse (compared to no intensity normalization, step 2) by an absolute value, though the only statistically significant drop in performance is Bias field correction on GBM dataset (82 mean Dice score (step 4.a) compared to 85 mean Dice score (step 4), $p_{\text{value}}=0.014$).

Second, for an attention-based model, the general trend stays the same, except for the  BGPD dataset and steps 4.a and b. Here, we observe a small but statistically significant increase in performance. We do not have a reasonable explanation for the effect (see \ref{AppendixA}), though we acknowledge that for all datasets, UNETR architecture results in worse performance compared to nnU-net. This might be the effect of a relatively small sample size, as transformer-based architectures require more training data. 

\subsection{Skull stripping} 

A brain mask application before training results in a moderate but statistically significant Dice score improvement for BGPD and LGG datasets (both nnU-net and UNETR) over the experiment without skull stripping, see Table \ref{tab-dices} 4 and 4.d. For the GBM dataset, the average segmentation quality is larger and the standard deviation is smaller in the experiment with skull stripping, though after a multiple comparisons correction these differences are not statistically significant ($p_{\text{value}} = 0.09$). 


\subsection{Volumetric errors}

In addition to the Dice scores, we provide the errors in tumour volume estimates for different preprocessing steps. 

We report the results for nnU-net in Table \ref{tab-volumetric}, as it performs better in our experiments. In terms of volume estimates, errors follow the same trend as with Dice scores: inter-subject alignment is always beneficial, image enhancement does not result in any improvements, and skull stripping systematically improves the quality. 


In volume estimates, skull stripping improves tumor segmentation quality for GBM and LGG datasets (step 4.d., Table \ref{tab-volumetric}) and does not for the BGPD dataset (for which there is a statistically significant improvement in terms of Dice scores). This result itself is aligned with the results in Dice scores, yet it is an argument for using clinically relevant metrics in addition to segmentation metrics. 

\begin{figure}
\includegraphics[width=\textwidth]{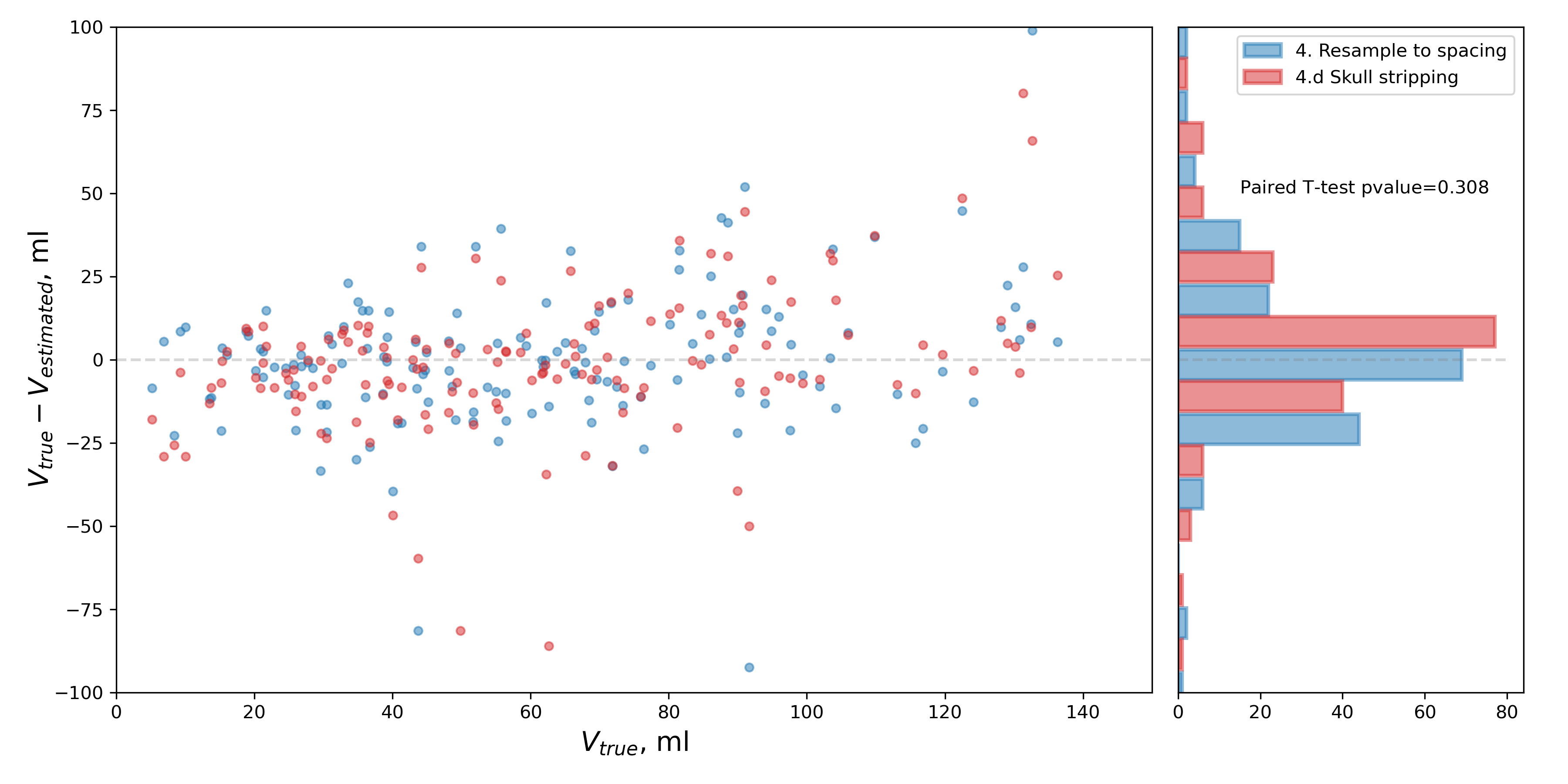}
\caption{The relation between tumor volume and its' estimated volume from predicted segmentation mask for two nnU-net experiments on BGPD: 4. Resample to spacing and 4.d Resample to spacing with skull stripping.}
\label{fig-volumes}
\end{figure}

Volumetric errors of model predictions on BGPD data with and without skull stripping are not statistically significant (MAE 26 (42) mL for skull stripping and MAE 28 (49) mL without skull stripping, $p_{\text{value}}= 0.308$), thus from a clinical perspective this additional step is not completely justified. Complete volumetric measurements for all experiments are provided in the \ref{AppendixA}.
Additionally, we check if error in volume estimate depends on tumor total size (Figure \ref{fig-volumes}), and do not observe any dependence.

\begin{table}
\centering
\caption{\label{tab-volumetric} Estimated MAE of model prediction from ground truth label for the three datasets: GBM
and LGG (WT label), BGPD (GTV label). Results are represented in mL, values are Mean (STD), the lower — the better. Volumes estimates are based on nnU-net. Arrows denote statistically significant difference ($p_{\text{value}} < 0.05$) compared to step 4. Resampling to spacing $\uparrow$ (increase), $\downarrow$ (decrease).}

    \begin{tabular}{lccc}
    \toprule
        Data preprocessing & $\text{GBM}$ & $\text{BGPD}$ & $\text{LGG}$ \\
        \toprule
        1.Inter-modality registration & $63\ (47)\downarrow$ & $55\ (62)\downarrow$ & $34\ (39)$\\
        \toprule
        2.Resampling to image size    & $15\ (14)$ & $28\ (43)$ & $23\ (24)$ \\
        3.Atlas registration          & $13\ (11)$ & $27\ (45)$ & $31\ (33)$\\ 
        \rowcolor{Gray}
        4.Resampling to spacing       & $14\ (13)$ & $28\ (49)$& $32\ (31)$\\ 
        \midrule
        4.a Bias field correction       & $15\ (15)$ & $27\ (49)$& $34\ (37)$  \\
        4.b Denoising                   & $13\ (13)$ & $27\ (47)$& $32\ (34)$\\
        4.c Histogram matching          & $14\ (12)$ & $27\ (47)$& $42\ (61)$ \\
        4.d Skull stripping             & $10\ (9)\uparrow$ & $26\ (42)$& $19\ (17)\uparrow$\\
        \bottomrule
    \end{tabular}   
\end{table}

\begin{figure}
     \centering
     \begin{subfigure}[b]{0.45\textwidth}
         \centering
         \includegraphics[width=\textwidth]{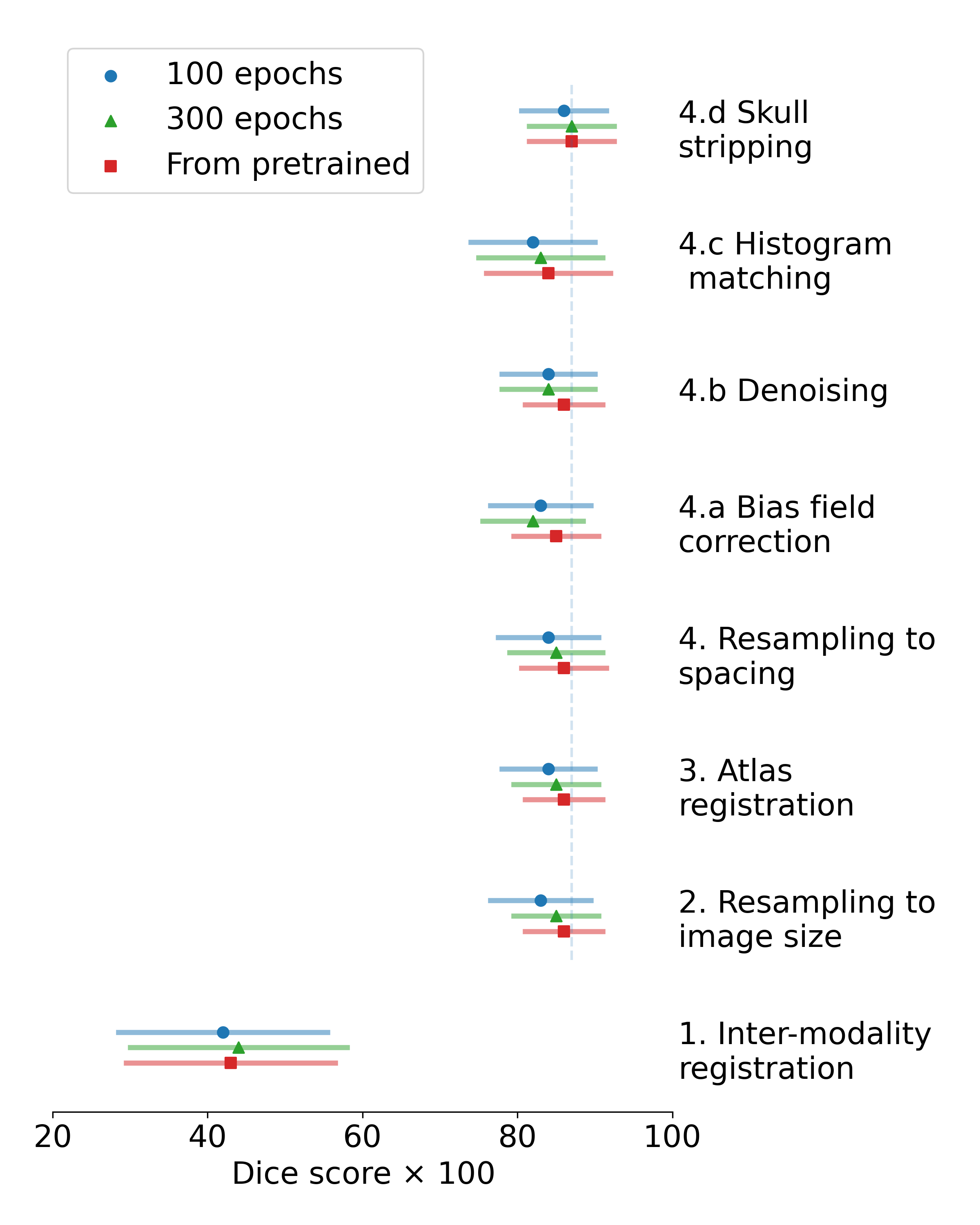}
         \caption{GBM dataset, WT label}
         \label{fig:gbm-errors}
     \end{subfigure}
     \hfill
     \begin{subfigure}[b]{0.45\textwidth}
         \centering
         \includegraphics[width=\textwidth]{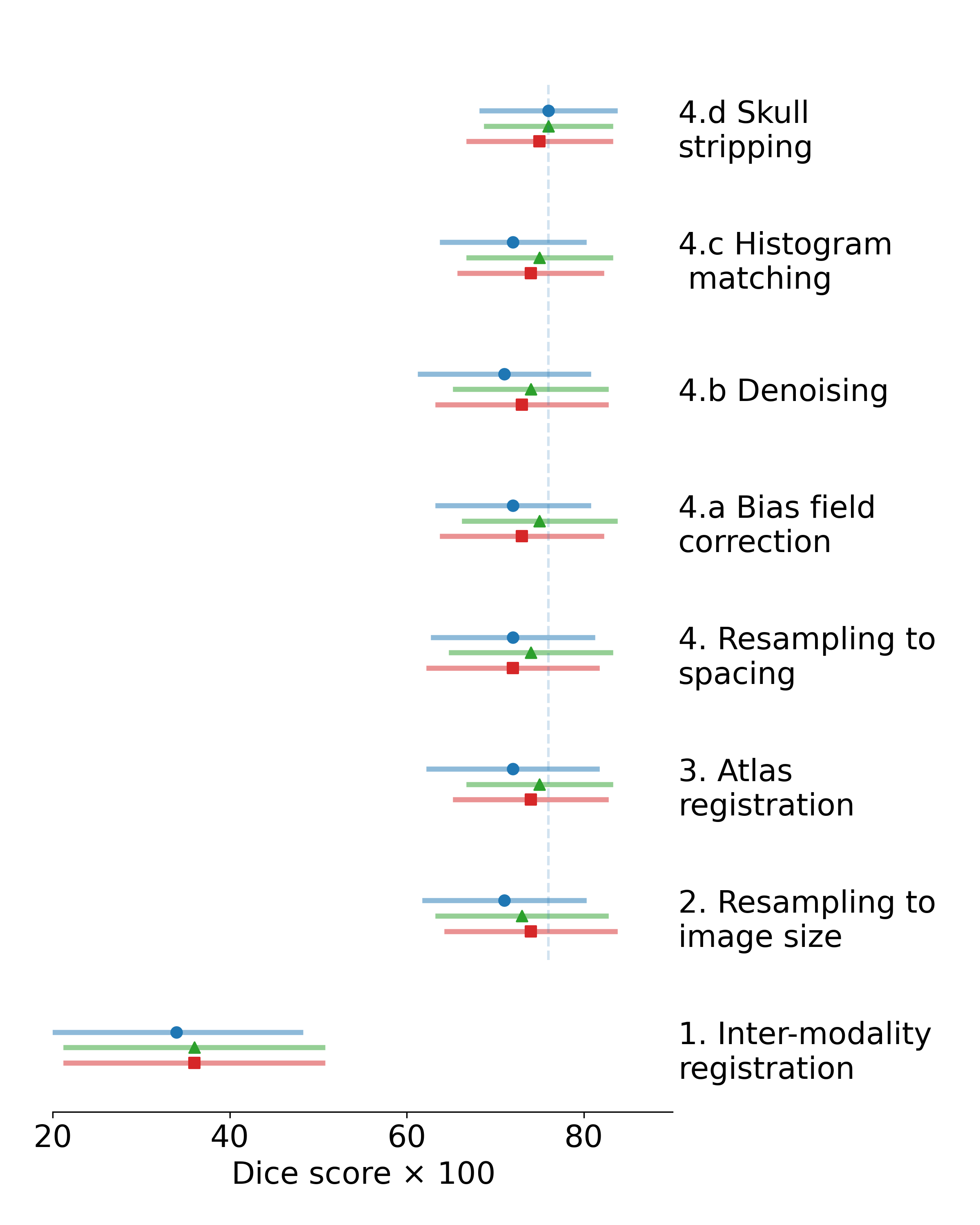}
         \caption{BGPD dataset, GTV label}
         \label{fig:bgpd-errors}
     \end{subfigure}
     \caption{nnU-net performance on GBM and BGPD datasets. Horizontal: segmentation accuracy presented in Dice scores from three fold cross-validation as $\text{Mean} \ (\text{STD})$ multiplied by 100. Vertical: preprocessing experiments 1-4.d for models trained with random weights initialization for 100 epochs (blue), 300 epochs (green) and fine-tuning for 100 epochs from pretrained weighs on another dataset (red).}
     \label{fig:error-bars}
\end{figure}

\subsection{Transfer learning.}
We test if MRI data preprocessing can facilitate model transfer from another dataset. In particular, we explore if model fine-tuning after training on preprocessed data is better than model fine-tuning on non-preprocessed data.

We repeat the main ablation study with nnU-net models pretrained on the GBM dataset for 300 epochs and fine-tune them on BGPD with the same data preprocessing for 100 epochs (and pretraining on BGPD with fine-tuning of GBM for corresponding experiments). We compare the Dice scores on three fold cross-validation for a model trained for 100 and 300 epochs from random weight initialization, and a model trained on 100 epochs from pretrained weights. The results are presented in Figure \ref{fig:error-bars}. 

First, we observe a definitive improvement in nnU-net performance on the GBM dataset with weight transfer and fine tuning for 100 epochs (red bars), compared to a training from scratch for 300 epochs (green bars), Figure \ref{fig:gbm-errors}. For the BGPD dataset, pretraining on the GBM sample results in better segmentation performance compared to training for 100 epochs (blue bars and red bars), but worse compared to training from scratch for 300 epochs (green bars), Figure \ref{fig:bgpd-errors}. This effect could be a consequence of the different sample sizes, as BGPD is almost two times larger than GBM — thus pretraining on the BGPD improves GBM segmentation, but not vice versa.

Second, we do not observe any differences in segmentation performance for either of the preprocessing steps on both datsets. For example, for the GBM dataset and preprocessing step 4.b Denoising, a model trained for 100 epochs results in 84 (12) Dice score (STD), the same if trained for 300 epochs, and 86 (10) Dice score if fine-tuned from BGPD. Yet, with the same data preprocessing on BGPD, we see a decrease in the mean Dice score with weight transfer from the GBM dataset to BGPD. Similarly, for step 4.d, the improvement of the segmentation quality on the GBM dataset is not observable in the BGPD sample. From these experiments we conclude that no preprocessing step among those studied improves model performance with weight transfer for both datasets. Numerical results depicted in Figure \ref{fig:error-bars} are accessible in \ref{AppendixD}.

Lastly, we perform an experiment with model fine-tuning from two large datasets BraTS2021, consisting of 2000 subjects \cite{baid2021rsna} and EGD dataset with 774 subjects.\cite{van2021erasmus}, with multiclass labels, similar to GBM and LGG ones. According to our results, model fine-tuning from a larger sample — the most exploited method of transfer learning \cite{ardalan2022transfer} — can be reached by longer training, irrespective of the size of the dataset for weight transfer, see \ref{AppendixC} Table \ref{transfer-big-datasets}.

\subsection{Our recommendations for brain MRI preprocessing for deep learning segmentation.}
The overall results suggest the following recommendations:
\begin{itemize}
    \item It's essential to align multi-modal MRI data between subjects for analysis, and even fast methods like image or voxel resizing yield comparable to atlas registration segmentation accuracy.
    \item Bias-field correction, denoising, and histogram matching are unnecessary in MRI segmentation pipelines based on UNet-like or UNETR architectures.
    \item Although skull stripping can improve segmentation performance, its impact on clinical measurements, such as differences in lesion volume estimates, is relatively small. Therefore, depending on the clinical task and the need for fast processing times, this step may not be necessary.
    \item Preprocessing MRI data does not help with transfer learning while fine-tuning models on other datasets. Moreover, there's almost no significant difference between fine-tuning models from other data and just doing longer training on the original sample.
\end{itemize}

\section{Conclusion}

We perform a rigorous ablation study of the most conventional preprocessing steps used in the analysis of brain MRI images, including atlas registration, voxel resampling and image resizing, histogram matching, bias-field correction, denoising and skull stripping. 

Although the image reprocessing steps might be useful for annotators and make distinct properties of the image more recognizable for the human eye, we show that only image alignment and voxel resampling are essential for accurate prediction with DNN models. We conclude that predictions after atlas registration do not significantly differ from ones with equal voxel resampling. We observe that bias-field correction, denoising, and histogram matching reduce data variance and do not affect DNN performance positively. We point out that skull stripping can lead to a measurable increase in accuracy and facilitate model convergence. On the other hand, brain extraction is very computationally expensive, and its incorporation into a pipeline does not affect clinically relevant volumetric measurements.

Thus we believe that skipping all steps excluding image alignment and voxel resampling from the brain MRI deep learning pipeline may reduce computational costs and improve reproducibility across studies. 

These recommendations will be especially relevant for MRI data preprocessing for semi-automated labeling with Segment Anything Model (SAM) and modifications \cite{ kirillov2023segany}. SAM is a vision-transformer based architecture, shown to be extremely useful for data annotation, yet still not surpassing the SOTA solutions for brain MRI data segmentation \cite{wu2023medical}. In the current work we define necessary preprocessing steps needed for MRI data annotation and further training, that will ensure reproducible across the studies and best segmentation accuracy.


\subsection{Work limitations} 

Our findings on data preprocessing strategies suggest that overall research reproducibility will benefit if one discards custom preprocessing steps, including different skull stripping, various implementations of bias field correction, denoising, etc. Yet we observed that the results of transfer learning and model training from scratch are strictly related to datasets' homogeneity and size. These effects could be different on  datests of thousands of images \cite{de2021effect}.

In the current study we focused on conventional brain MRI data preparation methods. The newly-developed methods of MRI harmonization, as multi site image harmonization by cumulative distribution function (CDF) alignment (MICA \cite{wrobel2020intensity}) or robust intensity distribution alignment (RIDA) \cite{sederevicius2022robust} could be outperforming the most conventional algorithms for histogram matching \cite{nyul2000new}. Advanced image intensity enhancement methods can be compared with the explored ones, i.e. with orthogonal moments \cite{da2022enhanced} for MR image enhancement. These analyses were outside the scope of the original study.

\subsection{Authors contributions}
Conceptualization and methodology, A.K., E.K. and P.D.; data pipeline organization and preprocessing pipeline E.K.; model training and analysis P.D.; interpretation of results, A.K. and A.D., S.Z., A.G.; writing original draft A.K., E.K. and P.D.; writing—review and editing, M.B., B.S. and A.D. All authors have read and agreed to the published version of the manuscript.


\subsection{Declaration of Competing Interest}
The authors declare that they have no known competing financial interests or personal relationships that could have appeared to influence the work reported in this paper.

\subsection{Acknowledgments}
This work was conducted in the Artificial Intelligence Research Institute in Moscow, Russia in collaboration with the National Medical Research Center for Neurosurgery and the Moscow Gamma Knife Center. The work of A.D., S.Z., A.G. and A.K. was supported by the Russian Foundation for Basic Research grant 18-29-01054. The work of P.D. and E.K. was supported by the Russian Science Foundation grant 21-71-10136.

\newpage
\bibliography{bibliography}

\newpage
\appendix

\section{The effect of image enhancement}
\label{AppendixA}

\begin{figure}[ht]
\includegraphics[width=\textwidth]{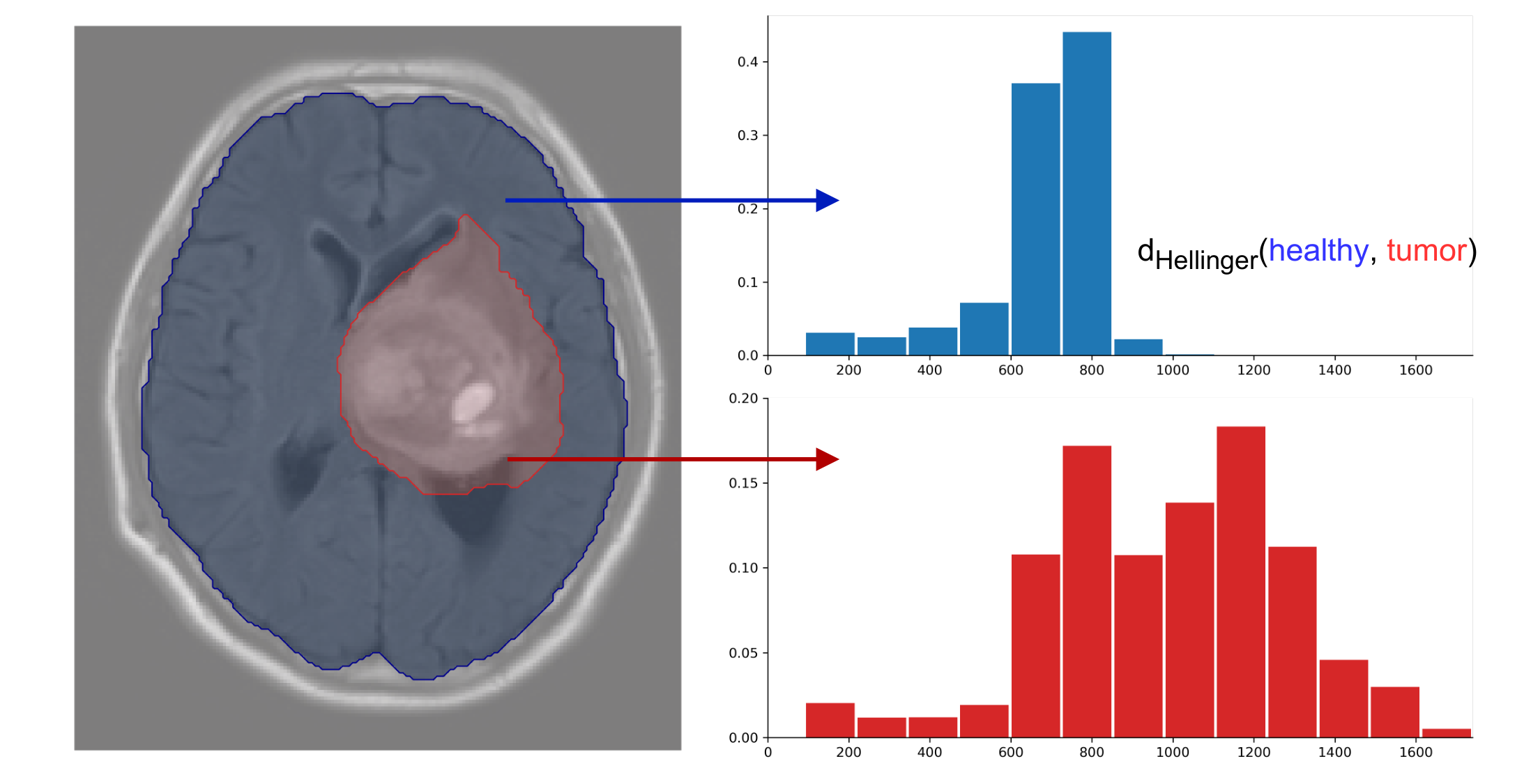}
\caption{Computing the distance between healthy tissue and tumor tissue regions.}
\label{fig-healthy-tumor}
\end{figure}

Why do image enhancement methods not facilitate segmentation? 

We attempt to explain why popular intensity normalization steps have questionable effect on segmentation performance. Our hypothesis is that while these steps equalize modalities appearance across the data, they also reduce the differences between voxels' intensities within each individual image. We compare intensity distribution for healthy brain voxels and voxels inside a tumor mask using Kullback-Leibler divergence (Table \ref{tab-kl}). We expect that if preprocessing steps increases KL divergence it should result in increased segmentation quality, and vice versa. In most of the cases this supposition holds. Two exceptions are Denoising for Tumor core segmentation and Histogram matching for Enhancing tumor segmentation.

We report KL values for histograms with fixed bin size (30 bins), and we test the Freedman-Diaconis heuristic (500-700 bins depending on subject) and Sturge’s rule (20-60 bins). Both approaches result in different absolute KL values but preserve a relative trend, without affecting the experiment conclusion.

In most cases, if the KL divergence decreases (differences between healthy and tumor tissue decrease), model performance decreases, too, and vice versa (steps 4.a-d in comparison to step 4, Table \ref{tab-kl}). In almost all cases, lower KL values corresponds to lower performance, e.g. for bias-field correction, the KL between healthy brain and WT tissue is equal to 0.47, compared to 0.61 of atlas registration, which coincides with a segmentation quality drop (from 86.4 for atlas registration to 84.9 for bias-field corrected data). On the contrary, for denoised data, the KL are either the same or slightly larger compared to atlas data: 0.63 vs 0.61 for WT; 4.16 vs 4.01 for TC and 7.11 vs 6.70 for ET, which completely coincides with segmentation performance. The only  comparison that does not follow this explanation is bf-correction for TC (it has a lower KL compared to atlas data, but slightly better segmentation quality).

\begin{table}[ht]
\caption{\label{tab-kl} KL divergence values (KL) and JS distance (JS) between intensity histograms for masked brain w/o tumor and tumor region. Lower values correspond to smaller differences in voxel intensities between healthy (brain mask without Whole Tumor(WT)  for GBM  and RT for BGPD sample) and tumor tissue. Stars represent statistically significant difference to 4. Resampling to spacing distances. }

\centering
\resizebox{\textwidth}{!}{
    \begin{tabular}{llccc}
        \toprule
         Data preprocessing  & GBM, KL & GBM, JS &  BGPD, KL &  BGPD, JS \\ 
        \midrule
        \rowcolor{Gray}
        4. Resampling to spacing & $0.830 (0.450)$  & $0.303(0.095)$ & $1.640(1.015)$  & $0.477(0.129)$   \\
        \midrule
        4.a Bias-field correction & $0.660 (0.300)$ & $0.259(0.087)$ & $1.145(0.723)$ &  $0.465(0.120)$   \\
        4.b Denoising & $0.870 (0.470)$ & $0.308(0.097)$ & $1.702(1.045)$ & $0.484(0.129)$ \\
        4.c Histogram matching & $0.770(0.470)$    & $0.304(0.095)$ & $1.650(1.029)$ & $0.478(0.130)$   \\   
        \bottomrule
    \end{tabular}}
\end{table}

\section{Model architectures and training} \label{AppendixC}

\begingroup
\setlength{\tabcolsep}{4pt}
\begin{table}[]
\centering
\caption{\label{tab-bgpd-dice} Hyperparameters for nnU-net and UNETR models. }
    \begin{tabular}{lcc}
    \toprule
        Parameter name & {nnU-net} & {UNETR} \\ 
        \toprule
        learning rate  &    $0.0003$ &  $0.0001$ \\ 
        weight decay & $0$ &  $0.00001$ \\
        momentum &   $0.99$ &  $0.99$\\
        patch size & $[128,128,128]$ & $[128,128,128]$\\
        batch size & $2$ & $2$ \\
        \bottomrule
    \end{tabular}
\end{table}
\endgroup

\textbf{nnU-net.} We use NVIDIA’s nn-Unet implementation for the BraTS2021 challenge\footnote{github.com/NVIDIA/DeepLearningExamples/tree/\\
ddbcd54056e8d1bc1c4d5a8ab34cb570ebea1947/PyTorch/Segmentation/nnUNet}. The following changes were applied by the authors on top of the default nnU-net architecture \cite{isensee2021nnu}: increasing the encoder depth to $7$, modifying the number of convolution channels to $[64, 96, 128, 192, 256, 384, 512]$, and using deep supervision — two additional output heads at the decoder levels. The model has multi-modal input of four modalities [T1, CT1, T2, FLAIR], plus a channel with one-hot encoding for foreground voxels, generated by image thresholding. We train nn-Unet with a patch size of \texttt{[128,128,128]} voxels and a batch size of two, a learning rate of $0.0003$ and a \texttt{Adam} optimizer with \texttt{momentum} of $0.99$.

\textbf{UNETR.} We use vision a transformer-based model UNETR \cite{Hatamizadeh2022UNETRTF} with an embedding size of $512$ for a 1D sequence of a 3D input of the same $5$ channels with patches of \texttt{[128,128,128]} voxels and the resolution of each patch equals 16. The model has $10$ heads and is trained with learning rate $0.0001$, with a \texttt{weight decay} of $0.00001$, and an \texttt{Adam} optimizer with \texttt{momentum} of $0.99$.

\textbf{Model optimization.} For the BGPD dataset we train the model to predict one class label. For the GBM dataset we train the model on three classes, according to the BraTS data labelling: WT, ET, TC. The model is then trained with the complex loss function. Each label class is optimized separately with the weighted sum of binary Cross-Entropy and the Dice loss (the trade-off weight value to 1 for both losses). The final complex loss function is optimized for a combination of class labels: the whole tumor (WT) (describes the union of the tumor core (ET), the non-enhancing part of the tumor core (NET) and the peritumoral edema ED), the tumor core (TC) (the union of the ET and NET), and the ET.

\begin{figure}[ht]
\includegraphics[width=\textwidth]{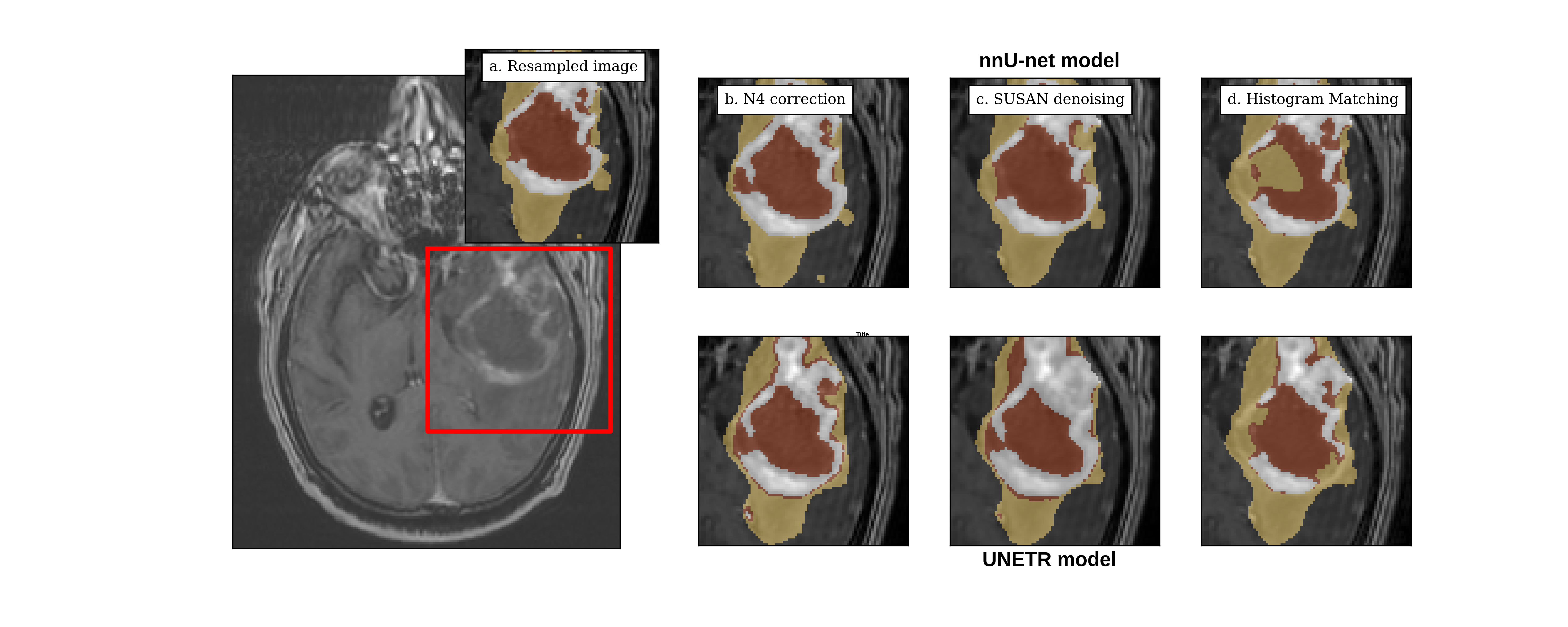}
\caption{Visualization of the two models' predictions with respect to the method of image intensity correction. The top raw (b-d) showing the image segmentation on 3 classes with nnUnet model (a)
an example CT1 MR image from GBM data after resampling to [1,1,1] voxel size; (b) N4
correction of the image; (c) SUSAN denoising of the image; (d) Histogram standartization; The bottom row showing tumor segmentation with UNETR model. Colors: ET - red, TC - white, WT - yellow.}
\label{fig-healthy-tumor}
\end{figure}

\section{Additional illustration of the results} \label{AppendixD}

In this section we show additional illustrations of the results not represented in the test's main body. In Tables \ref{tab-gbm-dice} and \ref{tab-lgg-dice} we show the results of other segmentation classes for multiclass tumor segmentation mask (WT, ET and TC) for the GBM dataset and (WT and ET) for the LGG dataset. 

In Table \ref{error-bars-table} we show the numerical representation of graphical results from Figure \ref{fig:error-bars}.


\begingroup
\setlength{\tabcolsep}{4pt} 
\begin{table}[]
\centering
\caption{\label{tab-gbm-dice}nnU-net and UNETR segmentation performance for three-class (Whole tumor, Enhancing tumor, Tumor core) segmentation on GBM dataset. Numbers are Dice scores mean (std) multiplied by 100. Trained for 300 epoch, columns 1 and 4 are duplicated from Table \ref{tab-dices}.}
\resizebox{\textwidth}{!}{
    \begin{tabular}{lcccccc}
        \toprule
        \multicolumn{1}{c}{} & \multicolumn{3}{c}{nnU-net} & \multicolumn{3}{c}{UNETR} \\
        \cmidrule(lr){2-4}\cmidrule(lr){5-7}
        Data preprocessing & {WT} & {ET} & {TC} & {WT} & {ET} & {TC} \\ 
        \toprule
        1. Inter-modality registration &  $44\ (28)\downarrow$ & $37\ (30)\downarrow$ & $30\ (27)\downarrow$ &$39\ (26)\downarrow$ &$32 \ (26)\downarrow$ & $28\ (24)\downarrow$\\ 
        \toprule
        2. Resampling to image size&      $85\ (11)$ & $80\ (18)$ & $74\ (19)$ &$82\ (12)$ &$75\ (18)$ &$72\ (19)$\\
        3. Atlas registration&            $85\ (11)$ & $80\ (17)$ & $72\ (20)$ &$82\ (13)$ &$74\ (18)$ &$69\ (21)$\\  
        \rowcolor{Gray}
        4. Resampling to spacing&         $85\ (12)$ & $80\ (16)$ & $73\ (19)$ &$83\ (14)$ &$75\ (20)$ &$71\ (22)$\\ 
        \midrule
        4.a Bias field correction&        $82\ (13)\downarrow$ & $80\ (17)$ & $72\ (19)$ &$80\ (13)$ &$75\ (20)$ &$71\ (21)$\\
        4.b Denoising&                    $84\ (12)$ & $80\ (17)$ &  $73\ (20)$& $83\ (13)$ &$75\ (20)$ &$70\ (22)$\\
        4.c Histogram matching&           $83\ (16)$ & $78\ (20)$ & $72\ (21)$ & $81\ (16)$ &$72\ (24)$ &$68\ (24)$\\ 
        4.d Skull stripping&              $87\ (11)$ & $82\ (16)$ & $76\ (19)$ &$85\ (11)$ &$79\ (16)$ &$75\ (19)$\\
        \bottomrule
    \end{tabular}}
\end{table}
\endgroup

\begingroup
\setlength{\tabcolsep}{4pt} 
\begin{table}[]
\centering
\caption{\label{tab-lgg-dice}nnU-net and UNETR segmentation performance for two-classes (Whole tumor, Enhancing tumor) segmentation on LGG dataset. Numbers are Dice scores $\text{mean} (\text{std})$ multiplied by 100, columns 1 and 3 are duplicated from Table \ref{tab-dices}.}
\resizebox{\textwidth}{!}{
    \begin{tabular}{lcccc}
        \toprule
        \multicolumn{1}{c}{} & \multicolumn{2}{c}{nnU-net} & \multicolumn{2}{c}{UNETR} \\
        \cmidrule(lr){2-3}\cmidrule(lr){4-5}
        Data preprocessing & {WT} & {ET} & {WT} & {ET}  \\ 
        \toprule
        1. Inter-modality registration &  $67\ (27)\downarrow$ & $47\ (29)$  &$66\ (23)$ &$49 \ (26)$ \\ 
        \toprule
        2. Resampling to image size&      $72\ (24)$ & $52\ (27)$  &$66\ (26)$ &$49\ (27)$ \\
        3. Atlas registration&            $71\ (25)$ & $52\ (29)$  &$67\ (25)$ &$49\ (27)$ \\  
        \rowcolor{Gray}
        4. Resampling to spacing&         $70\ (25)$ & $48\ (28)$  &$67\ (23)$ &$50\ (25)$ \\ 
        \midrule
        4.a Bias field correction&        $67\ (25)$ & $45\ (29)$ &$62\ (22)$ &$45\ (25)$ \\
        4.b Denoising&                    $70\ (26)$ & $51\ (30)$ & $65\ (25)$ &$48\ (27)$ \\
        4.c Histogram matching&           $68\ (26)$ & $49\ (29)$  & $63\ (26)$ &$45\ (28)$ \\ 
        4.d Skull stripping&              $77\ (21)\uparrow$ & $54\ (25)\uparrow$ &$75\ (19)\uparrow$ &$54\ (25)\uparrow$ \\
        \bottomrule
    \end{tabular}}
\end{table}
\endgroup

\begingroup
\setlength{\tabcolsep}{4pt} 
\begin{table}
\centering
\caption{\label{error-bars-table}nnU-net performance on GBM and BGPD datasets, numerical representation of Figure \ref{fig:error-bars}. Segmentation accuracy presented in Dice scores from three fold cross-validation as $\text{Mean} \ (\text{STD})$ multiplied by 100. Models trained with random weights initialization for 100 epochs, 300 epochs and fine-tuning for 100 epochs from pretrained weighs on other dataset (BGPD-GBM, GBM-BGPD finetune). Columns 2 and 5 are duplicated from Table \ref{tab-dices}.}
\resizebox{\textwidth}{!}{
    \begin{tabular}{lccacca}
        \toprule
        \multicolumn{1}{c}{} & \multicolumn{3}{c}{GBM} & \multicolumn{3}{c}{BGPD} \\
        \cmidrule(lr){2-4}\cmidrule(lr){5-7}
        Data preprocessing              & {100 epochs}     & {300 epochs }      & {BGPD-GBM}      & {100 epochs}     & {300 epochs}    & {GBM-BGPD} \\ 
        \toprule
        1. Inter-modality registration &  $42\ (27)$ & $44\ (28)$ & $43\ (27)$ &$34\ (28)$ &$36\ (29)$ & $36\ (29)$\\ 
        \toprule
        2. Resampling to image size&      $83\ (13)$ & $85\ (10)$ & $86\ (10)$ &$71\ (18)$ &$73\ (19)$ &$74\ (19)$\\
        3. Atlas registration&            $84\ (12)$ & $85\ (11)$ & $86\ (10)\uparrow$ &$72\ (19)$ &$75\ (16)$ &$72\ (17)$\\  
        4. Resampling to spacing&         $84\ (13)$ & $85\ (12)$ & $86\ (11)$ &$72\ (18)$ &$74\ (18)$ &$72\ (19)$\\ 
        4.a Bias field correction&        $83\ (13)$ & $82\ (13)$ & $85\ (11)\uparrow$ &$72\ (17)$ &$75\ (17)$ &$73\ (18)$\\
        4.b Denoising&                    $84\ (12)$ & $84\ (12)$ &  $86\ (10)\uparrow$ & $71\ (19)$ &$74\ (17)$ &$73\ (19)$\\
        4.c Histogram matching&           $82\ (16)$ & $83\ (16)$ & $84\ (26)$ & $72\ (16)$ & $75\ (16)$ &$74\ (16)$\\ 
        4.d Skull stripping&              $86\ (11)$ & $87\ (11)$ & $87\ (11)$ & $76\ (15)$ & $76\ (14)$ &$75\ (16)$\\
        \bottomrule
    \end{tabular}}
\end{table}

\begingroup
\setlength{\tabcolsep}{4pt} 
\begin{table}
\centering
\caption{\label{transfer-big-datasets}nnU-net performance with and w/o model transfer from large datasets (BraTS2021 \cite{baid2021rsna} and EGD \cite{van2021erasmus}. Segmentation accuracy presented in Dice scores from three fold cross-validation as $\text{Mean} \ (\text{STD})$ multiplied by 100. Models were trained for 300 epochs and fine-tuned for 100 epochs from pretrained weights on the larger dataset  (BraTS-BGPD, EGD-GBM, BraTS-EGD finetune). Columns 1 and 3 are duplicated from Table \ref{tab-dices}.}
\resizebox{\textwidth}{!}{
    \begin{tabular}{lcacaca}
        \toprule

        Data preprocessing         
        & \multicolumn{1}{p{2cm}}{\centering BGPD, \\ 300 epochs}
        & \multicolumn{1}{p{2cm}}{\centering BraTS-BGPD, \\ Finetune}
        & \multicolumn{1}{p{2cm}}{\centering GBM, \\ 300 epochs}
        & \multicolumn{1}{p{2cm}}{\centering EGD-GBM, \\ Finetune}
        & \multicolumn{1}{p{2cm}}{\centering EGD, \\ 300 epochs}
        & \multicolumn{1}{p{2cm}}{\centering BraTS-EGD, \\ Finetune} \\ 
        \toprule
        3. Atlas registration&            $75\ (17)$ & $74\ (20)$ & $85\ (12)$ &$84\ (11)$ &$ - $ &$ - $\\  
        \midrule
        4.d Skull stripping&              $76\ (14)$ & $75\ (17)$ & $87\ (11)$ & $86\ (10)$ & $ 83\ (12) $ &$ 83\ (12)$\\
        \bottomrule
    \end{tabular}}
\end{table}

\end{document}